\documentclass[useAMS,usenatbib,usegraphicx]{mn2e}

\title[UV-to-optical spectrum of primordial galaxies]{Rest-frame
ultraviolet-to-optical spectral characteristics of extremely metal-poor
and metal-free galaxies}
\author[A. K. Inoue]{Akio K. Inoue$^{1}$\thanks{E-mail: 
akinoue@las.osaka-sandai.ac.jp}\\
$^{1}$College of General Education, Osaka Sangyo University, 
3-1-1, Nakagaito, Daito, Osaka 574-8530, Japan}
\begin{document}

\date{}

\pagerange{\pageref{firstpage}--\pageref{lastpage}} \pubyear{2009}

\maketitle

\label{firstpage}

\begin{abstract}
Finding the first generation of galaxies in the early Universe is the
 greatest step forward for understanding galaxy formation and
 evolution. For strategic survey of such galaxies and interpretation of
 the obtained data, this paper presents an ultraviolet-to-optical
 spectral model of galaxies with a great care of the nebular
 emission. In particular, we present a machine-readable table of
 intensities of 119 nebular emission lines from Ly$\alpha$ to the
 rest-frame 1 $\mu$m as a function of metallicity from zero to the Solar
 one. Based on the spectral model, we present criteria of equivalent
 widths of Ly$\alpha$, He {\sc ii} $\lambda1640$, H$\alpha$, H$\beta$, 
 [O {\sc iii}] $\lambda5007$ to select extremely metal-poor and
 metal-free galaxies although these criteria have uncertainty caused by
 the Lyman continuum escape fraction and the star formation duration. We
 also present criteria of broad-band colours which will be useful
 to select candidates for spectroscopic follow-up from drop-out
 galaxies. We propose the line intensity ratio of [O {\sc iii}]
 $\lambda5007$ to H$\beta$ $<0.1$ as the most robust criterion for
 $<1/1000$ of the Solar metallicity. This ratio of a galaxy with a few
 $M_\odot$ yr$^{-1}$ at $z\sim8$ is detectable by spectroscopy with
 the James Webb Space Telescope within a reasonable exposure time.
\end{abstract}


\section{Introduction}

Understanding galaxy formation and evolution is one of the most
important issues in the modern astronomy. Finding the first generation
of galaxies is the largest step to solve the question, while it is also
an open question what the first generation of galaxies is. Galaxies in
the very early Universe should be metal-poor or may be even metal-free. 
Therefore, the most metal-poor galaxies would be the first generation.

The galaxy with the lowest known metallicity is I Zw 18, a blue compact
dwarf galaxy in the local Universe. The measured gas metallicity (to be
precisely oxygen abundance) is about 1/50 of the Sun \citep{izo97}. If
we adopt the classical Solar metallicity $Z_\odot=0.02$ \citep{and89},
it corresponds to $Z=0.0004$. At $z\sim2$--3, Lyman break galaxies
(LBGs) selected by the so-called drop-out technique have
$Z\sim0.002$--0.01 ($=1/10$--1/2 $Z_\odot$) 
\citep[e.g.,][]{pet01,erb06,man09,erb10}. The metallicity
measurements for galaxies selected by the strong Ly$\alpha$ emission
line, Ly$\alpha$ emitters (LAEs), are still very rare because of the
difficulty of spectroscopy for these faint galaxies. \cite{fin11} report
$Z\la0.004$ ($=1/5$ $Z_\odot$) at $z\sim2$. These measurements are made
by some strong nebular emission lines, so that the measured metallicity
is that in the ionized gas not in the stellar atmosphere.

There are hundreds of stars whose metallicity in the atmosphere is
found to be extremely low $Z<1\times10^{-5}$ $(=1/2000~Z_\odot)$ in the
halo of the Galaxy \citep[e.g.,][]{bee05}. The mass of such extremely
metal-poor (EMP) stars in the present epoch is small ($\sim1$ $M_\odot$), 
while their high-mass counter part should exist in the past \citep{kom07}.
Indeed, there are some possible signatures suggesting that LBGs and LAEs
at $z\ga3$ contain massive EMP stars or even metal-free stars, the
so-called Population III (Pop III) stars, of a non-negligible fraction in
their stellar mass \citep{mal02,jim06,bou10a,ino11}.

To find the first generation of galaxies, we are pushing out the
redshift frontier. The current record of redshift measured by
spectroscopy is $z=8.55$ of UDFy-38135539 \citep{leh10}. Like this
object, the brand-new Hubble Space Telescope/Wide Field Camera 3
(HST/WFC3) imaging enabled us to select LBGs at $z\ga7$ and even at
$z\sim10$ \citep[e.g.,][]{bou10b,bou11}.
These highest-$z$ galaxies may contain more EMP or Pop III stars than
lower-$z$ LBGs. The very blue ultraviolet (UV) colours of the HST/WFC3
LBGs suggest such a possibility \citep{bou10a} \citep[but see][]{dun11}. 
Future telescopes such as 30-m class extremely large telescopes and
James Webb Space Telescope (JWST) will further push out the redshift
frontier by imaging observations.

Spectroscopy is finally required to measure metallicity of galaxies as
well as the precise redshift of them. However, it is time-consuming
because of the faintness of the target galaxies. Thus, any
pre-selections for spectroscopy will be useful. We will start from a
sample selected by the standard drop-out technique first. Then, it may
be useful if we can select EMP or even metal-free candidates only with
imaging data, in particular, broad-band photometric colours. This paper
presents such a method.

In this paper, we will present a spectral model of EMP or even metal-free
galaxies with a great care of the nebular emission (both lines and
continuum). Young starburst galaxies emit strong Lyman continuum
(i.e. hydrogen ionizing continuum with wavelength less than 912 \AA;
hereafter LyC). Thus, the nebular emission is a very important spectral
component \citep{zac08,sch09,sch10,ino10,ono10,rai10}. This model will
be useful to discuss the physical nature of very high-$z$ LBGs with
broad-band data (i.e. so-called spectral energy distribution [SED]
fit) as well as to select EMP or metal-free galaxies for follow-up
spectroscopy.

In section 2, we will describe the modelling of nebular emission and SED
of galaxies. In section 3, we will present the resultant spectra from UV
to optical in the rest-frame for various metallicities and present
equivalent widths of emission lines as a function of metallicity and 
star formation duration. In section 4, we compare our model with the
observed broad-band colours of $z\sim7$--8 galaxies and discuss how we
find EMP and Pop III galaxies in the future. In the final section, we
will present a summary of this paper.

This paper follows a standard $\Lambda$CDM cosmology with 
$\Omega_{\rm M}=0.3$, $\Omega_{\Lambda}=0.7$, and $h=0.7$.
All the magnitude is described in the AB system.

\section{Model}

\subsection{Stellar spectra}

The SED of pure stellar populations depends on metallicity $Z$, initial
mass function (IMF), star formation history, and age. Here, we assume a
Salpeter IMF \citep{sal55} with 1--100 $M_\odot$ and a constant star 
formation with a duration of 1, 10, 100, or 500 Myr. The metallicities
considered are $Z=0.02$ ($=Z_\odot$), 0.008, 0.004, 0.0004,
$1\times10^{-5}$ (EMP), $1\times10^{-7}$, and 0 (Pop III). SEDs of the
former 4 cases are generated by the population synthesis code {\sc
starburst99} version 5.1 \citep{lei99}. Those for the latter 3 cases are
taken from \cite{sch02,sch03}.

\subsection{Nebular emission}

\subsubsection{Lines}

Emission lines of hydrogen, helium, and some other major elements are
taken into account. Based on a large grid of photo-ionization models by
using {\sc cloudy} 08.00 \citep{fer98}, we have obtained emissivities 
of emission lines relative to H$\beta$. Table~1 shows the parameter
space of nebulae explored by the {\sc cloudy} calculations in this paper: 7
metallicities $Z$, 5 ionization parameters $U$, and 5 hydrogen number
densities $n_{\rm H}$. The set of $Z$ is the same as that of the stellar
spectra adopted (\S2.1). For each case of $Z$, we input the
corresponding stellar spectrum into the code. Since there are 4 cases of
the duration of star formation for each $Z$, the total number of
the model calculated is 700 ($=4\times7\times5\times5$). The considered
$U$ and $n_{\rm H}$ are typical ranges of Galactic and extra-galactic H
{\sc ii} regions \citep[e.g.,][]{ost06} and the validity of the values
will be confirmed by a comparison with observations for some strong
lines later (Fig.~1). We assume the plane-parallel geometry for all the
calculations and assume the abundance of elements to be the Solar one as
the default in the code. We consider two cases with and without dust in
ionized nebulae (see \S2.2.3 for details). The electron temperatures in
the nebulae are calculated in the code with energy balance between
heating and cooling rates. Table~2 shows average temperatures for each
$Z$ cases which are used to calculate H$\beta$ luminosity by equations
(1) and (2) later and nebular continuum emission in \S2.2.2. The higher
temperatures in dusty cases are probably due to photoelectric heating by
dust.

\begin{table}
 \caption[]{Nebular parameters explored in this paper.}
 \setlength{\tabcolsep}{3pt}
 \footnotesize
 \begin{minipage}{\linewidth}
  \begin{tabular}{lc}
   \hline
   Parameter & Values \\
   \hline
   $\log_{10}(Z/Z_\odot)$ & $\infty$, $-5.3$, $-3.3$, $-1.7$, $-0.7$,
       $-0.4$, 0.0\\
   $\log_{10}U$ & $-3.0$, $-2.5$, $-2.0$, $-1.5$, $-1.0$\\
   $\log_{10}(n_{\rm H}/{\rm cm^{-3}})$ & 0.0, 0.5, 1.0, 1.5, 2.0\\
   \hline
  \end{tabular}
 \end{minipage}
\end{table}%

\begin{table}
 \caption[]{Average electron temperatures in the nebulae for various
 metallicities.}
 \setlength{\tabcolsep}{3pt}
 \footnotesize
 \begin{minipage}{\linewidth}
  \begin{tabular}{lcc}
   \hline
   & no dust & dusty\\
   $\log_{10}(Z/Z_\odot)$ & $T_{\rm e}$ (kK) & $T_{\rm e}$ (kK)\\
   \hline
   $\infty$ & 20 & ---\\
   $-5.3$ & 20 & 20\\
   $-3.3$ & 20 & 20\\
   $-1.7$ & 18 & 18\\
   $-0.7$ & 12 & 13\\
   $-0.4$ & 9.5 & 10\\
   0.0 & 5.4 & 6.3\\
   \hline
  \end{tabular}
 \end{minipage}
\end{table}%

\begin{figure*}
 \begin{center}
  \includegraphics[width=10cm]{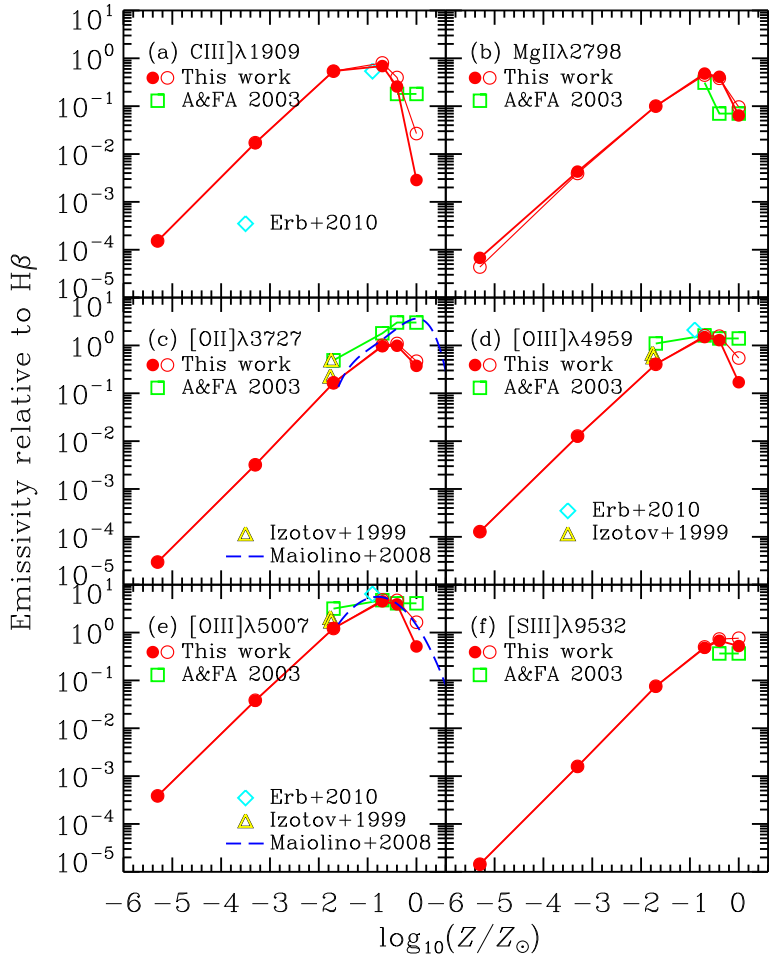}
 \end{center}
 \caption{Emissivities relative to H$\beta$ of six strongest metal
 emission lines as a function of metallicity. The circles are results
 calculated with the code {\sc cloudy} 08.00 (Ferland 1998) in this
 work: no dust model (filled) and dusty model (open). The squares
 are empirical results by Anders \& Fritze-v.\ Alvensleben (2003). 
 The triangles are the observations of I Zw 18 (northwest and southeast
 components) measured by Izotov et al.~(1999). The diamonds are
 observations of a $z=2.3$ galaxy measured by Erb et al.~(2010). The
 dashed lines are empirical results by Nagao et al.~(2006) and updated
 by Maiolino et al.~(2008).}
\end{figure*}

We select 119 emission lines between Ly$\alpha$ and 1 \micron\ in the
rest-frame: H Ly$\alpha$, Balmer and Paschen series, and He and other
elements' lines which have an emissivity more than about 1\% of H$\beta$
in at least one case calculated (except for H lines, for which we adopt
all the lines output from the {\sc cloudy}: up to $n=24$). The list of
the lines is presented in Appendix.

For each stellar spectrum (i.e. each $Z$ and star formation duration),
we calculated 25 set of ($U$, $n_{\rm H}$). Among the 25 cases, line
emissivities relative to H$\beta$ vary within about an order of
magnitude for metal lines but the standard deviation relative to the
average is about 5--25\%. The variations of H and He lines are much
smaller and the relative standard deviations are typically $<5$\%. For
calculations in later sections, we adopt average line emissivities
obtained for each set of $Z$ and star formation duration. We have also
found that the line emissivities are very stable against the change of
star formation duration if it is larger than 10 Myr. This is
probably because the saturation of the LyC luminosity and spectrum after
$\ga10$ Myr star formation. In Appendix, we present machine-readable
tables of line emissivities relative to H$\beta$ as a function of only
$Z$ (i.e. averaged over 3 star formation durations of 10, 100, and 500
Myr) which may be useful for future calculations by readers. On the
other hand, in this paper, we adopt the line emissivities depending each
star formation duration.

Figure~1 shows the relative emissivities of 6 strongest metal emission
lines as a function of metallicity $Z$ (ones presented in Appendix,
i.e. averaged over the star formation durations). The circles are
theoretical results of this paper by {\sc cloudy} 08.00: no dust case
(filled) and dusty case (open). The squares and dashed lines are
empirical relations compiled by \cite{and03} and \cite{mai08} (see also 
\citealt{nag06}), respectively. The triangles and diamonds are
observations of a local galaxy, I Zw 18, \citep{izo99} and a $z=2.3$
galaxy \citep{erb10}. We find a good overall agreement between
ours and theirs but an order of magnitude difference is found for some
cases. This indicates a large uncertainty of the line emissivities which
depend on the nebular physical parameters such as $U$ and 
$n_{\rm H}$. Indeed, \cite{nag06} suggest $Z$ dependence of $U$ to
explain their empirical relations. Despite such uncertainties and
difficulties, our emissivity of the strongest line, [O {\sc iii}]
$\lambda5007$, excellently agrees with that of \cite{nag06} who compiled
the largest sample of galaxies distributed over the widest range of
$Z$ and updated by \cite{mai08}. As pointed out by literature
\citep{zac08,sch09,sch10} and also seen in later sections, this [O {\sc
iii}] emission line has the largest effect on broad-band colours and
other metal lines do not affect the colours significantly. Therefore, we
consider that using our theoretical line emissivities (or our choice of
$U$ and $n_{\rm H}$) is justified and a large uncertainty of line
emissivities does not degrade our conclusions based on the  [O {\sc
iii}] line.

To obtain the luminosity of the emission lines based on the relative
emissivities, we need the luminosity of H$\beta$. We assume the
following expression:
\begin{equation}
 L_{\rm H\beta} = \frac{\gamma_{\rm H\beta}(T_{\rm e})
  Q_*(1-f_{\rm esc}-f_{\rm dust})}
  {\alpha_{\rm B}(T_{\rm e})
  +\alpha_1(T_{\rm e})(f_{\rm esc}+f_{\rm dust})}\,,
\end{equation}
where $\gamma_{\rm H\beta}$ is the H$\beta$ emission coefficient, $Q_*$
is the stellar production rate of LyC photons, $f_{\rm esc}$ is
the escape fraction of the photons, $f_{\rm dust}$ is the fraction of
the photons absorbed by dust within the ionized gas, $\alpha_{\rm B}$ is
the Case B recombination rate, and $\alpha_1$ is the recombination rate
to the ground state. The denominator is usually described as just
$\alpha_{\rm B}$. However, our expression is correct when the nebular
LyC escapes from the nebulae and is absorbed by dust with the same
probability of $f_{\rm esc}$ and $f_{\rm dust}$ as the stellar LyC 
and the ionization equilibrium is established in the nebulae
\citep{ino10}. The fractions of $f_{\rm esc}$ and $f_{\rm dust}$ is
assumed to be independent of wavelength, which is valid for clumpy
nebulae as discussed in \cite{ino10}. The LyC escape should be taken
into account especially for $z\ga3$ because a significant escape was
detected at $z\simeq3$ \citep{sha06,iwa09} and such galactic ionizing
radiation probably caused the cosmic reionization. On the other hand, we
omit the LyC absorption in this paper (i.e. $f_{\rm dust}=0$) because
the effect is very small for low-metallicity cases, according to a
discussion in \S2.2.3.

In $\gamma_{\rm H\beta}$, we have omit the density dependence because we
only consider densities smaller than 10$^2$ cm$^{-3}$ which is well
below a criterion for the small density limit ($<10^4$ cm$^{-3}$;
\citealt{ost06}). The electron temperature $T_{\rm e}$ dependence of
$\gamma_{\rm H\beta}$ can be approximated to 
\begin{equation}
 \gamma_{\rm H\beta} = 1.23 \times 10^{-25}
  \left(\frac{T_{\rm e}}{10^4~{\rm K}}\right)^{-0.9}
  ~{\rm erg~s^{-1}~cm^3}\,,
\end{equation}
which is obtained from Table~B.5 in \cite{dop03} for the density of
$10^2$ cm$^{-3}$. The uncertainty is less than 4\% between 5,000 K and
30,000 K. This emission coefficient is obtained with the Case B
assumption \citep{dop03}. Recently \cite{rai10} have shown that H$\beta$
luminosity is excellently predicted by the Case B even for extremely
metal-poor cases. However, they also show that the Case B prediction
underestimates Ly$\alpha$ luminosity. On the other hand, our prediction
of Ly$\alpha$ is based on the ratio of Ly$\alpha$ to H$\beta$ obtained
from the {\sc cloudy}, in which the effect discussed by \cite{rai10} is
already taken into account. 

\begin{figure}
 \begin{center}
  \includegraphics[width=6cm]{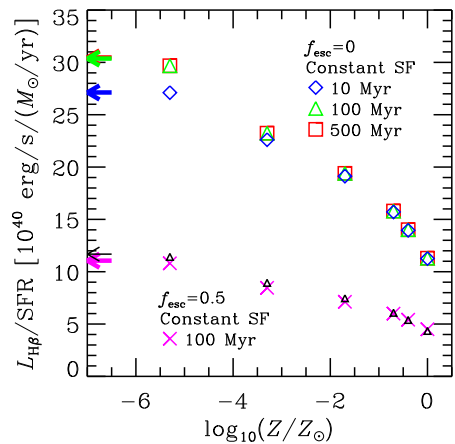}
 \end{center}
 \caption{H$\beta$ luminosity per unit star formation rate as a function
 of metallicity. The diamonds, triangles, and squares are the cases of
 the LyC escape fraction $f_{\rm esc}=0$ and the constant star formation
 duration of 10, 100, and 500 Myr. The crosses are the case of 
 $f_{\rm esc}=0.5$ and 100 Myr. The small triangles are the same
 case but obtained by the $f_{\rm esc}$ scaling of equation (3). The
 arrows at the left edge of the panel are the metal-free cases. The LyC
 absorption in nebulae is neglected (i.e. $f_{\rm dust}=0$).}
\end{figure}

Figure~2 shows H$\beta$ luminosity as a function of metallicity of
the stellar population. The luminosity is normalized by the star
formation rate. The H$\beta$ luminosity increases when metallicity
decreases because the ionizing photon production rate $Q_*$ is larger
when metallicity is lower. The dependence of star formation
duration is weak or absent when the duration is longer than 10--100 Myr
depending on the metallicity. The dependence of the escape fraction
$f_{\rm esc}$ found in equation (1) can be approximated to 
\begin{equation}
 L_{\rm H\beta}(f_{\rm esc})\approx
  L_{\rm H\beta}(0)\times\frac{1-f_{\rm esc}}{1+0.6f_{\rm esc}}\,,
\end{equation}
where the factor 0.6 in the denominator comes from 
$\alpha_1/\alpha_{\rm B}$ for $T_{\rm e}=10^4$ K. Note that the scaling
is not just $1-f_{\rm esc}$ as usually assumed when the nebular LyC can
escape. This scaling is very nice as shown by the comparison of the
crosses with the triangles in Figure~2.

\subsubsection{Continuum}

As the nebular continuum, we consider bound-free, free-free, and two
photon emissions of hydrogen. Any helium continuum emission is not taken
into account in this paper. While this simplification underestimates the
nebular continuum emission, the helium continuum is not important very
much because it is weak and negligible in fact \citep[see e.g.,][]{ino10}. 
Continua of metal elements are not taken into account, either. The
luminosity density of hydrogen nebular continuum at the frequency $\nu$
is given by the very similar form to equation (1) as
\begin{equation}
 L_\nu = \frac{\gamma_\nu(T_{\rm e})
  Q_*(1-f_{\rm esc}-f_{\rm dust})}
  {\alpha_{\rm B}(T_{\rm e})
  +\alpha_1(T_{\rm e})(f_{\rm esc}+f_{\rm dust})}\,,
\end{equation}
where $\gamma_\nu$ is the emission coefficient of the continuum. The
volume emissivity $\gamma_\nu$ and the recombination rates $\alpha_{\rm
B}$ and $\alpha_1$ are calculated as described in \cite{ino10}. 
\cite{rai10} have shown that the Case B assumption, which is adopted in
\cite{ino10}, underestimates the nebular two-photon emission for the
stellar effective temperature of $>5\times10^4$ K and the density of
$<10^2$ cm$^{-3}$. This point causes an underestimation of the nebular
continuum for $\lambda \la 0.2$ $\mu$m, while the effect is negligible
for $\lambda>0.2$ $\mu$m where the nebular continuum is dominated by the
bound-free emission.

\begin{figure*}
 \begin{center}
  \includegraphics[width=11cm]{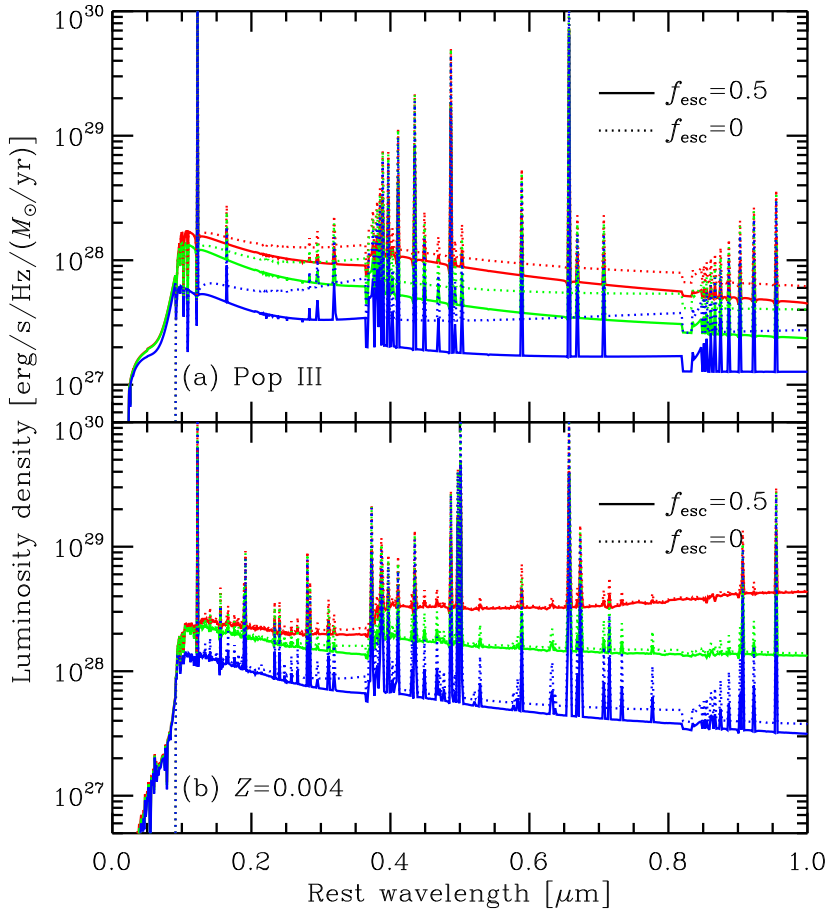}
 \end{center}
 \caption{Rest-frame model spectra of (a) Pop III ($Z=0$) galaxies and
 (b) moderate sub-solar metallicity ($Z=0.004$: $\log_{10}(Z/Z_\odot)=-0.7$)
 galaxies. In each panel, the lines correspond to 10, 100, and 500 Myr
 of constant star formation (bottom to top). LyC escape fractions
 $f_{\rm esc}=0.5$ and 0 are assumed for the solid and the dotted lines,
 respectively. The emission line width is assumed to be 300 km s$^{-1}$
 as an example. No attenuation by dust and IGM is not included.}
\end{figure*}

\subsubsection{Dust}

There are two effects of dust; one is attenuation (or extinction) of
radiation by dust in the interstellar medium (mainly in the 
{\it outside} of ionized nebulae). This effect is often described with
the Calzetti law \citep{cal00}, and we do so in \S4, although some
studies have suggested a different attenuation law in high-$z$ 
\citep[e.g.,][]{sia09,gal10}. 

The other effect is LyC absorption by dust {\it within} ionized
nebulae (i.e. $f_{\rm dust}$ in eqs.~[1] and [3]). Dust competes with
hydrogen to absorb LyC in nebulae. Indeed, a half of LyC is directly
absorbed by dust in H {\sc ii} regions in the Milky Way and some local
group galaxies \citep{ino01a,ino01b}. The amount of LyC absorption
depends metallicity and dust-to-gas ratio \citep{ino01b}. To see this
effect, we ran {\sc cloudy} with the interstellar dust of the code
(i.e. Milky Way dust). The amount of dust was simply assumed to be
scaled by metallicity. We also assumed that dust was well mixed with
gas, while a central dustless cavity is likely to exist in nebulae
\citep{ino02}. The uniform dust distribution may overestimate the effect
of LyC absorption. As found from Figure~1, the dust effect on emission
lines results in larger emissivities relative to H$\beta$ in most cases,
because LyC absorption reduces H$\beta$ emissivity. However,
this effect is not very large, especially for low-metallicity
cases. Therefore, we omit LyC absorption (i.e. $f_{\rm dust}=0$) and
adopt only no dust cases in the following sections.

\section{Result}

\subsection{Rest-frame UV-to-optical spectrum}

Figure~3 shows resultant model spectra in the rest-frame for two cases
of $Z$: (a) $Z=0$ (Pop III) and (b) $Z=0.004$ ($\log_{10}(Z/Z_\odot)=-0.7$). 
A constant star formation rate is assumed and the duration of star
formation is 10, 100, or 500 Myr (bottom to top in each panel). The
LyC escape fraction $f_{\rm esc}$ is assumed to be 0.5 (solid) or 0
(dotted) for both stellar and nebular continua. We do not consider any
attenuation by dust and IGM in this figure. Note that the vertical axis
is normalized by a unit star formation rate.

In both cases of $Z$, many H recombination lines such as Ly$\alpha$,
H$\alpha$, H$\beta$ are highly visible. In addition, some He lines such
as He {\sc ii} $\lambda1640$ and He {\sc i} $\lambda5876$ are also
remarkable in the Pop III case. In the $Z=0.004$ case, many metal
emission lines appear. Especially, metal emission lines shown in
Figure~1 such as [O {\sc iii}] $\lambda4959/\lambda5007$ lines are very
strong. These lines affect even the broad-band colours as found 
later.\footnote{A spectral dip just longward of the Paschen limit (8204
\AA) is not real but an artifact caused by the lack of Paschen series
lines higher than $n=24$.}

\begin{figure}
 \begin{center}
  \includegraphics[width=6cm]{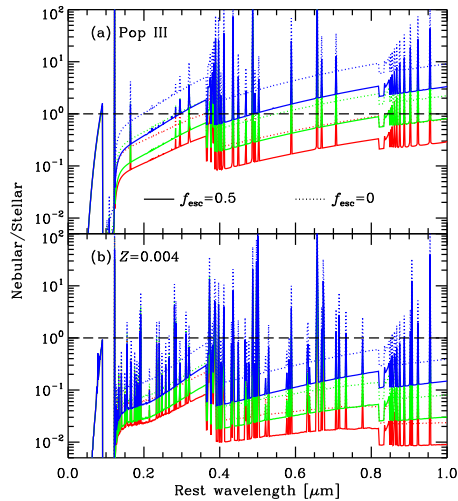}
 \end{center}
 \caption{Rest-frame spectra of luminosity density ratio of nebular
 emission to stellar emission: (a) Pop III and (b) moderate sub-solar
 metallicity ($Z=0.004$: $\log_{10}(Z/Z_\odot)=-0.7$). In each panel,
 the solid lines correspond to the LyC escape fraction $f_{\rm esc}=0.5$
 and the dotted lines correspond to $f_{\rm esc}=0$. For each type of
 the lines, the top to bottom correspond to 10, 100, and 500 Myr of
 constant star formation. The dashed horizontal lines indicate the ratio
 to be unity (i.e. equal contribution from stellar and nebular
 emissions).}
\end{figure}

The nebular contribution to the continuum is more significant in the
$Z=0$ case than in the $Z=0.004$ case. The Balmer jump in the bound-free
continuum is easily recognised in the $Z=0$ and the duration of 10 Myr
case. This point is shown more clearly in Figure~4 which shows spectra
of luminosity density ratio of nebular to stellar emissions. The lines
top to bottom correspond to the duration of 10, 100, and 500 Myr. As the
duration of star formation increases, the ratio decreases. In addition,
we see a stronger nebular contribution for a longer wavelength.

There are three spectral jump in the nebular continuum within the
wavelength range shown: Lyman, Balmer, and Paschen jump of the
bound-free continuum. In particular, the Balmer jump at 3646 \AA\ is
strong and can be found in the total (stellar+nebular) spectra shown in
Figure~3. For example, the Pop III case with 10 Myr duration in Figure~3
shows a factor of $\sim2$ Balmer jump. Note that the `Balmer jump' is a
sudden decrease of the continuum level towards longer wavelength. This
is opposite to the `Balmer break' which is a sudden increase of the
continuum towards longer wavelength found in spectra of older stellar
populations. For example, the 100 and 500 Myr cases in Figure~3 (b) show
a prominent Balmer break.

\begin{figure}
 \begin{center}
  \includegraphics[width=6cm]{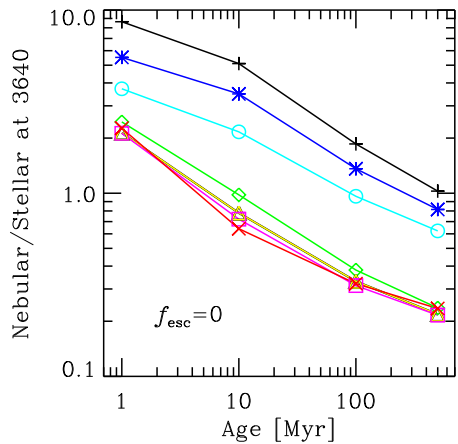}
 \end{center}
 \caption{Luminosity density ratio of nebular to stellar emissions at
 3640 \AA, very close to the Balmer limit (3646 \AA), as a function of
 the duration of star formation. There are 7 metallicity cases:
 $\log_{10}(Z/Z_\odot)=\infty$ (plus), $-5.3$ (asterisk), $-3.3$
 (circle), $-1.7$ (diamond), $-0.7$ (triangle), $-0.4$ (square), and 0.0
 (cross). No escape of the LyC is assumed ($f_{\rm esc}=0$).}
\end{figure}

The strength of the Balmer jump depends on the duration of star
formation and metallicity. This point is shown in Figure~5 more clearly:
the nebular-to-stellar ratio at just shortward of the Balmer limit as a
function of the duration. The LyC escape fraction of $f_{\rm esc}=0$ is
assumed in Figure~5, and thus, the ratio shown is the maximum. The
$f_{\rm esc}$ dependence of the nebular-to-stellar ratio is simply the
same as equation (3). For all the 7 cases of metallicity, the
nebular contribution, or the strength of the Balmer jump, monotonically
decreases as the duration increases. The nebular contribution also
decreases for higher metallicity. For $\ga10$ Myr duration, only EMP
($\log_{10}(Z/Z_\odot)\leq-3.3$) or metal-free cases exceed the ratio of
unity, i.e. stronger nebular emission. Thus, a prominent Balmer jump
appears only for these cases, while normally sub-solar metallicity
(i.e. $\log_{10}(Z/Z_\odot)\geq-1.7$) cases expect a weaker Balmer
jump. Therefore, the strong Balmer jump can be a signature of EMP or Pop
III stellar populations.

In Figure~4, we can find another strong jump at the Lyman limit. This
feature is also caused by H bound-free emission and appears only when
$f_{\rm esc}>0$. As proposed by \cite{ino10}, this Lyman jump (or
`bump') can be also useful as a signature of EMP or Pop III stellar
populations because the jump becomes strong enough only when EMP or
metal-free \citep{ino10}. \cite{ino11} showed that strong LyC emission
detected by \cite{iwa09} from $z\simeq3.1$ LAEs can be attributed to the
Lyman jump by EMP or Pop III stars. However, this feature is easily
obscured by IGM attenuation at $z\ga4$ \citep{ino10} and we cannot use
it to find EMP or Pop III galaxies at very high-$z$.

\subsection{Equivalent width of emission lines}

The equivalent width (EW) of an emission line is defined as 
\begin{equation}
 {\rm EW}_{\rm line} \equiv \frac{L_{\rm line}(f_{\rm esc})}
  {L^*_{\lambda_{\rm line}} + L^{\rm neb}_{\lambda_{\rm line}}(f_{\rm esc})}\,,
\end{equation}
where $L_{\rm line}$ is the line luminosity, $L^*_{\lambda_{\rm line}}$
is the stellar luminosity density at the line wavelength 
$\lambda_{\rm line}$, and $L^{\rm neb}_{\lambda_{\rm line}}$ is the
nebular continuum luminosity density at $\lambda_{\rm line}$. As
explicitly expressed in the equation, $L_{\rm line}$ and 
$L^{\rm neb}_{\lambda_{\rm line}}$ depend on the LyC escape
fraction $f_{\rm esc}$. The dependence of these two terms is the same as
equation (3). Thus, the $f_{\rm esc}$ dependence of EW has two extreme
cases: (a) the same form as equation (3) when 
$L^*_{\lambda_{\rm line}} \gg L^{\rm neb}_{\lambda_{\rm line}}$ and 
(b) independent of $f_{\rm esc}$ when 
$L^*_{\lambda_{\rm line}} \ll L^{\rm neb}_{\lambda_{\rm line}}$.
Therefore, the $f_{\rm esc}$ dependence of EW is not very simple when
the nebular continuum is relatively strong. Such cases happen at the
rest-frame optical when the star formation duration is very short
($\sim1$ Myr) or when metallicity is extremely low ($Z<10^{-5}$) as
shown in the previous subsection.

\subsubsection{Ly$\alpha$}

\begin{figure}
 \begin{center}
  \includegraphics[width=6cm]{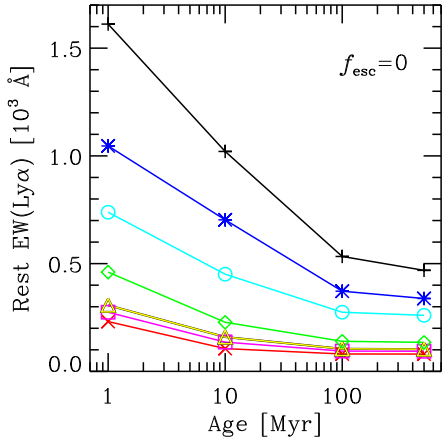}
 \end{center}
 \caption{Rest-frame equivalent width of Ly$\alpha$ emission line as a
 function of the duration of star formation. The symbols correspond to 7
 metallicities same as Fig.~5. No escape of the LyC is assumed 
 ($f_{\rm esc}=0$).}
\end{figure}

Figure~6 shows the rest-frame EW of Ly$\alpha$ as a function of the
duration of star formation. The LyC escape fraction $f_{\rm esc}=0$ is
assumed. We can scale the EW for other $f_{\rm esc}$ by the 
$f_{\rm esc}$ dependence same as equation (3) excellently well because
the nebular continuum at 1216 \AA\ is weak enough. We see that the EW
decreases monotonically as the duration becomes longer. We also see that
the EW is smaller as metallicity is higher. These are already reported
in literature \citep[e.g.,][]{sch02,sch03,rai10}. Note that our 
EWs are quantitatively very consistent with those with the same 
IMF and a similar duration in \cite{rai10}.

We expect the maximum EW for galaxies with $Z\ge 0.0004$ 
($\log_{10}(Z/Z_\odot)=-1.7$) to be 460 \AA,
or to be 230 \AA\ if we consider only the age larger than 10 Myr (Note
that the probability to observe a galaxy as young as 1 Myr is generally
small because of its short time). Therefore, we may conclude that
galaxies with an EW $>230$ \AA\ (or $>460$ \AA\ for more conservative)
are EMP or Pop III \citep[see also][]{mal02}. However, Ly$\alpha$
transfer in the interstellar medium (ISM) is complex, and sometimes, it
enhances the EW \citep{neu91}. Thus, a galaxy with higher metallicity
may have a boosted EW which exceeds the criterion, and then, may be
identified as an EMP/Pop III galaxy. On the other hand, it is worth
noting that the EW becomes smaller than the criterion even for
EMP/Pop III galaxies if $f_{\rm esc}>0$. For example, we find that the
EW is 65 \AA\ for $Z=0$ and 10 Myr constant star formation when 
$f_{\rm esc}=0.9$. Therefore, Ly$\alpha$ EW is intriguing but we need
other signatures simultaneously to conclude a galaxy to be EMP/Pop III.

\subsubsection{He {\sc ii}}

\begin{figure}
 \begin{center}
  \includegraphics[width=6cm]{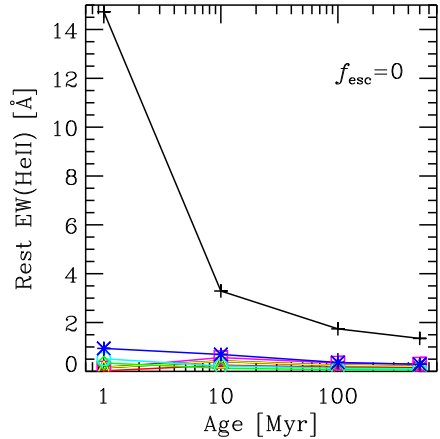}
 \end{center}
 \caption{Same as Fig.~6, but for He {\sc ii} $\lambda$1640.}
\end{figure}

Figure~7 shows the rest-frame EW of He {\sc ii} $\lambda$1640. We see
that the EW $>1$ \AA\ is realized only when metallicity is zero
(i.e. Pop III). Therefore, this emission line is proposed as the
signature of Pop III stars \citep[e.g.,][]{sch02,sch03}. However, this
line may be too weak to be detected, and what is worse, the EW becomes
even smaller than those in Figure~7 if $f_{\rm esc}>0$. The scaling of
$f_{\rm esc}$ in equation (3) is good for $\ga10$ Myr cases because the
nebular continuum contribution is not very large at the wavelength of
the line for the cases. Even if we detect the He {\sc ii} line, there is
another issue that the line becomes observable from Wolf-Rayet stars
with normal metallicity. In fact, the line has been detected from
$z=2$--3 LBGs but its origin is attributed to the stellar winds of these
stars \citep{sha06,erb10}. Therefore, using the He {\sc ii} line needs
more careful considerations, for example, about the line width
\citep{erb10}.

\subsubsection{H$\alpha$ and H$\beta$}

\begin{figure}
 \begin{center}
  \includegraphics[width=6cm]{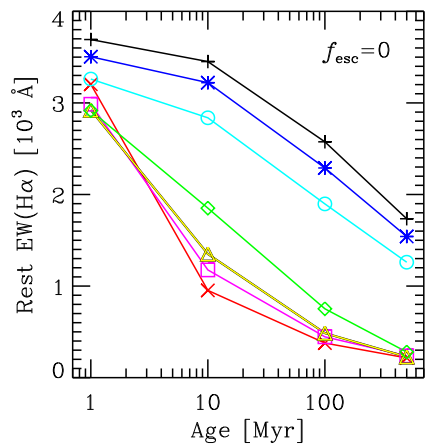}
 \end{center}
 \caption{Same as Fig.~6, but for H$\alpha$.}
\end{figure}

\begin{figure}
 \begin{center}
  \includegraphics[width=6cm]{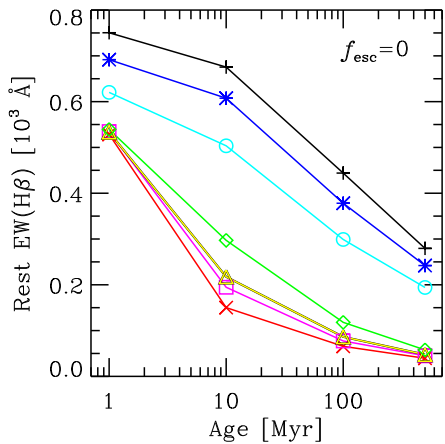}
 \end{center}
 \caption{Same as Fig.~6, but for H$\beta$.}
\end{figure}

Figures~8 and 9 show the rest-frame EWs of H$\alpha$ and H$\beta$,
respectively. These two cases show qualitatively similar results: the EW
becomes smaller as metallicity is higher or the duration is longer.
For $\sim10$--100 Myr durations, we can separate EMP and Pop III cases
from other higher metallicity cases by criteria, for example,
EW(H$\alpha$) $>1900$ \AA\ and EW(H$\beta$) $>300$ \AA. If we wish to
remove very young galaxies with higher metallicities, the criteria
should be EW(H$\alpha$) $>3200$ \AA\ and EW(H$\beta$) $>540$ \AA. Unlike
Ly$\alpha$, H$\alpha$ and H$\beta$ photons do not undergo resonant
scattering. Therefore, there are any contaminants of galaxies with
higher metallicities in the sample selected by these criteria, except
for AGNs. In this sense, the H$\alpha$ and H$\beta$ EWs are promising
tool to find primordial galaxies in the near future.

Note that we will miss EMP and Pop III galaxies with $f_{\rm esc}>0$ by
these criteria. The $f_{\rm esc}$ dependence of the EWs is not simple
because the nebular continuum significantly contributes to the total
continuum at H$\alpha$ and H$\beta$, especially for the EMP and Pop III
cases. The scaling of equation (3) overestimates the EWs by a factor of
2 or more, unfortunately. We need full calculations in order to obtain
the EWs for other $f_{\rm esc}$ with higher accuracy.

\subsubsection{[O {\sc iii}]}

\begin{figure}
 \begin{center}
  \includegraphics[width=6cm]{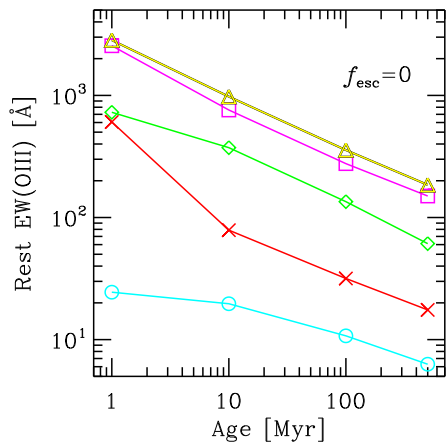}
 \end{center}
 \caption{Same as Fig.~6, but for [O {\sc iii}] $\lambda$5007.}
\end{figure}

Figure~10 shows the rest-frame EW of [O {\sc iii}] $\lambda$5007 which
is the strongest metal emission line in our expected spectra. As shown
in Figure~1, the metallicity dependence of this line is not monotonous.
The line strength becomes maximum at around $\log_{10}(Z/Z_\odot)=-0.7$.
For higher or lower than the metallicity, the EW becomes lower. We
expect that EMP and Pop III galaxies have the EW $<20$ \AA. If we
consider $f_{\rm esc}>0$, the EW becomes even lower. Thus, all EMP and
Pop III galaxies satisfy this criterion. On the other hand, the EW for
a higher metallicity will also satisfy the criterion if $f_{\rm esc}>0$.
This will be contaminant. In addition, very old quiescent galaxies with
higher metallicities, if they exist at high-$z$, are also contaminant.

\section{Discussion}

\begin{figure}
 \begin{center}
  \includegraphics[width=7cm]{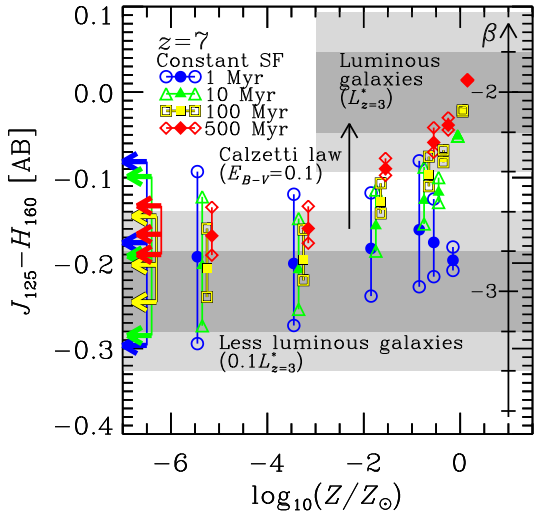}
 \end{center}
 \caption{HST/WFC3 $J_{125}-H_{160}$ colours of $z=7$ galaxies for seven
 metallicities. Each metallicity, there are four cases of the star
 formation duration: 1 Myr (circle), 10 Myr (triangle), 100 Myr
 (square), and 500 Myr (diamond). The metal-free case are indicated by
 arrows at the left edge of the panel. The four models of each
 metallicity are horizontally shifted each other for the display
 purpose. Each model has three cases of the LyC escape fraction 
 $f_{\rm esc}$: the upper open symbols for $f_{\rm esc}=0$, the
 middle filled symbols for $f_{\rm esc}=0.5$, and the lower open symbols
 for $f_{\rm esc}=1$. The vertical arrow is the dust reddening vector based on
 Calzetti et al.~(2000), which is likely to be applied for higher
 metallicity cases. At the right edge of the panel, the UV slope
 $\beta$ based on the conversion by equation (1) in Bouwens et
 al.~(2010a) is shown. The two shaded regions are the observed ranges of
 $\beta$ for luminous and less luminous galaxies at $z\sim7$ reported by
 Bouwens et al.~(2010a); the thick and thin shades indicate the ranges
 within 1-$\sigma$ and 2-$\sigma$ uncertainties, respectively.}
\end{figure}

\begin{figure}
 \begin{center}
  \includegraphics[width=7cm]{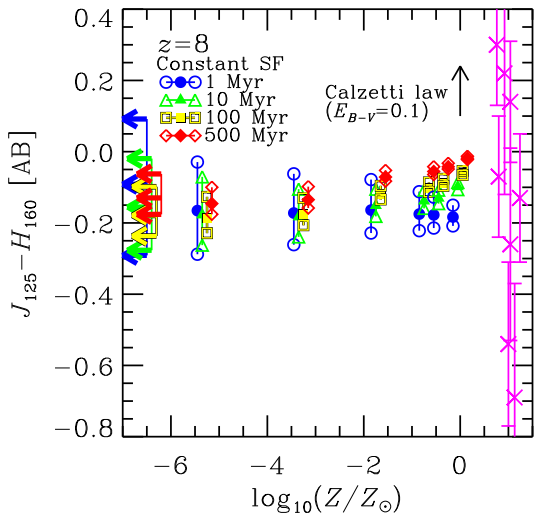}
 \end{center}
 \caption{Same as Fig.~11, but for $z=8$ galaxies. Note the much wider
 range of the colours than Fig.~11. The crosses with vertical error-bars
 near the right edge of the panel are the eight robust sample of
 $z\sim8$ galaxies compiled by Taniguchi et al.~(2010). Their
 positions along the horizontal axis are meaningless and just for the
 sake of showing.}
\end{figure}

\subsection{Rest-UV colours of high-$z$ galaxies with HST/WFC3}

\subsubsection{$z\sim7$}

\cite{bou10a} have reported very blue UV colours of $z\sim7$ less
luminous LBGs found in the ultra-deep survey with HST/WFC3, while
\cite{dun11} recently challenged their finding. The argument by
\cite{bou10a} is that the UV slope $\beta=-3$ found in the LBGs with 
$M_{\rm UV}=-19$ to $-18$ AB indicates extremely low metallicity as 
$\log_{10}(Z/Z_\odot)\leq-3.3$ and large escape fraction of the LyC
as $f_{\rm esc}>0.3$. Let us examine this argument with our
model spectra.

Figure~11 shows a comparison of our model with the observed UV slope (or
$J-H$ colour) by \cite{bou10a}. For each metallicity, we consider four
cases of the star formation duration (1, 10, 100, and 500 Myr) and three
cases of the LyC escape fraction (0, 0.5, and 1). We find
that almost all the EMP and Pop III cases with 
$\log_{10}(Z/Z_\odot)\leq-3.3$ reside within a 2-$\sigma$ range of the
observed $\beta=-3.0\pm0.2$ of the less luminous LBGs. The most
plausible duration and $f_{\rm esc}$ among the EMP and Pop III cases are
the duration of $\la100$ Myr and $f_{\rm esc}=0.5$--1. For the
metallicity $\log_{10}(Z/Z_\odot)\geq-1.7$, the observed $\beta$ is
reproduced only when the duration is $\sim1$ Myr, while the 10 Myr cases
of $\log_{10}(Z/Z_\odot)=-1.7$ and $-0.7$ still remain within the upper 
2-$\sigma$ range if $f_{\rm esc}>0.5$. It is unlikely to observe
galaxies with $\sim1$ Myr age because of the short time to observe them.
Furthermore, the higher metallicity cases are probably reddened by dust,
which makes more difficult to reproduce the observed $\beta$ with such
higher metallicities. Therefore, we agree with the \cite{bou10a}'s
argument if their very blue $\beta$ is real.

On the other hand, \cite{bou10a} have reported the UV slope
$\beta=-2.0\pm0.2$ for luminous LBGs ($M_{\rm UV}=-21$ to $-20$ AB) at
$z\sim7$. Without dust reddening, we need the Solar metallicity and the
star formation duration of $\ga100$ Myr to reach the $\beta$. If we
apply a small amount (e.g., $E_{B-V}=0.1$) of the dust reddening to the
model, lower metallicity cases can reach the $\beta$. However, the EMP
and Pop III cases are unlikely to have dust enough and to reach the
observed $\beta$. Therefore, the luminous LBGs at $z\sim7$ probably have
metallicity larger than $0.01Z_\odot$.

\subsubsection{$z\sim8$}

\cite{tan10} have presented a robust sample of galaxies at $z\sim8$
detected with HST/WFC3 \citep{bou10b,bun10,mcl10,yan10,fin10}. Let us
compare the UV colours of the galaxies with our model. Figure~12 shows
the comparison. However, a large uncertainty and variance on the
observed colour make it difficult for us to derive any implication from
the comparison. Thus, we comment just one thing about the bluest two
objects which are much bluer than our bluest model (i.e. EMP and Pop III
cases with $f_{\rm esc}=1$). This may indicate the presence of a strong
Ly$\alpha$ emission line in the $J_{125}$ band of the two objects 
\citep{tan10}. Indeed, the line enters into the $J_{125}$ band if the
object is located at $z\ge8.1$. Note that the colour at $z=8$ shown in
Figure~12 does not include the effect of the line.

\subsection{Rest-optical colours of high-$z$ galaxies with JWST/NIRCAM}

\begin{figure*}
 \begin{center}
  \includegraphics[width=11cm]{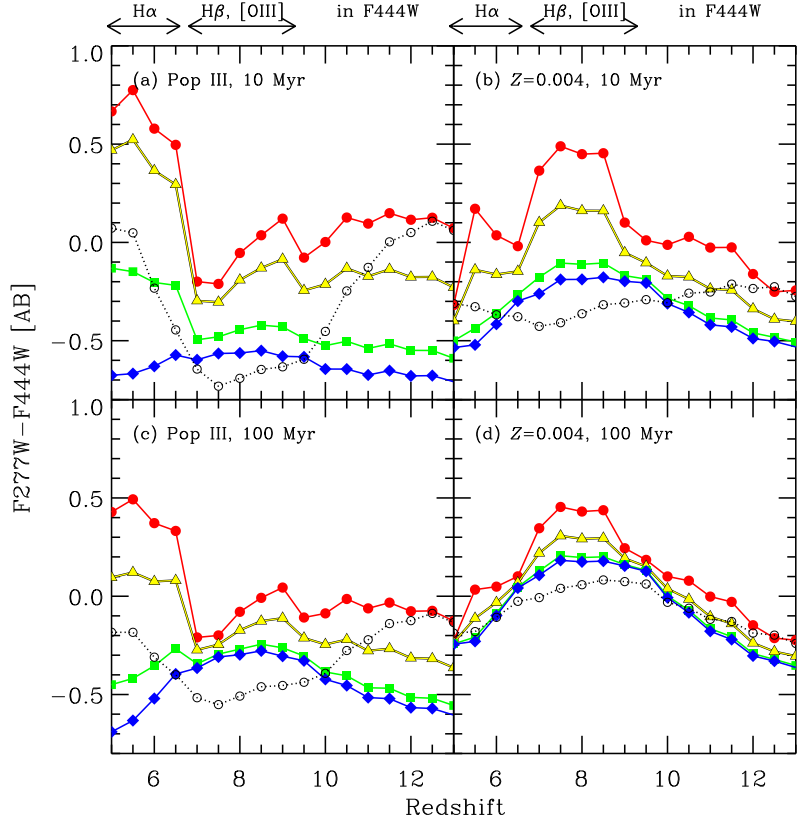}
 \end{center}
 \caption{JWST/NIRCAM F277W$-$F444W colours of galaxies as a function of
 redshift: (a) Pop III ($Z=0$) galaxies with 10 Myr constant star
 formation, (b) $Z=0.004$ ($\log_{10}(Z/Z_\odot)=-0.7$) galaxies with 10
 Myr star formation, (c) Pop III case but 100 Myr star formation, and (d)
 $Z=0.004$ case but 100 Myr star formation. Four LyC escape fractions
 $f_{\rm esc}$ are assumed: $f_{\rm esc}=0$ for the filled 
 circles, $f_{\rm esc}=0.5$ for the triangles, $f_{\rm esc}=0.9$ for the
 squares, and $f_{\rm esc}=1$ (i.e. pure stellar spectrum) for the
 diamonds. The open circles are the case with $f_{\rm esc}=0$ but no
 emission lines (i.e. stellar+nebular continua).}
\end{figure*}

As discussed in section 3.1, there are the Balmer jump and some strong
emission lines in the rest-frame optical. The observed wavelength of
these features from high-$z$ becomes near- and mid-infrared where
ground-based observations are difficult. The forthcoming JWST will cover
the wavelength from the space. Let us examine the effect of the optical
features on broad-band colours observed with the Near Infrared Camera
(NIRCAM) on the JWST. In particular, we focus on F244W$-$F444W colour
because the Balmer jump at $z\sim8$ comes between the two
bands.\footnote{The wavelength coverages of the two broad-bands taken
from http://ircamera.as.arizona.edu/nircam/.}

Figure~13 shows the redshift dependence of the colour. Although we show
only four combinations of metallicity and star formation duration as
indicated in the panels, the trend of other cases is similar to these
cases. We find that the nebular emission makes the colour redder
than the pure stellar one. The degree of the reddening increases as the
LyC escape fraction $f_{\rm esc}$ decreases. This reddening is mainly
caused by the emission lines like H$\alpha$, H$\beta$, and [O {\sc iii}]
as found by the comparison between the cases with and without emission
lines which are indicated by the filled and open circles, respectively.
In fact, the nebular continuum makes the colour bluer than the stellar
one for $z\sim8$ galaxies because of the Balmer jump. However, the
reddening by strong H$\beta$ and [O {\sc iii}] emission lines in the
F444W band overcomes the bluing by the Balmer jump. We also find that
for a longer star formation duration, the reddening by the nebular
emission is smaller. In the same time, the stellar colour is redder
because of the Balmer break of older stars. As a result, the colour with
the nebular emission (with the both of lines and continuum) does not
depend on the duration strongly.

The difference of metallicity appears in the colour. At $z\sim6$,
H$\alpha$ is in the F444W band, and then, F277W$-$F444W $\ga0$ for the
Pop III cases with $f_{\rm esc}\ga0.5$ but the $Z=0.004$ cases are
almost always F277W$-$F444W $\la0$. At $z\sim8$, the $Z=0.004$ cases
show F277W$-$F444W $\ga0$ due to [O {\sc iii}] in the F444W band, but
the Pop III cases show F277W$-$F444W $\la0$ thanks to the lack of the 
[O {\sc iii}] line. Figure~14 shows the metallicity dependence of the
colour for $z\sim8$ galaxies more in detail. As found from the figure,
the EMP and Pop III cases with $\log_{10}(Z/Z_\odot)\leq-3.3$ expect 
F277W$-$F444W $\la0$. On the other hand, the higher metallicity cases
expect F277W$-$F444W $\ga0$, except for the cases of very young
($\la10$ Myr) and $f_{\rm esc}\ga0.5$. Note that the colour of the
higher metallicity cases is the lower limit. These galaxies probably
have dust which reddens the colour as indicated by the upper arrow in
the panel. Therefore, a colour criterion of F277W$-$F444W $<0$ may be
useful to select EMP and Pop III candidates from galaxies at $z\sim8$.

\begin{figure}
 \begin{center}
  \includegraphics[width=7cm]{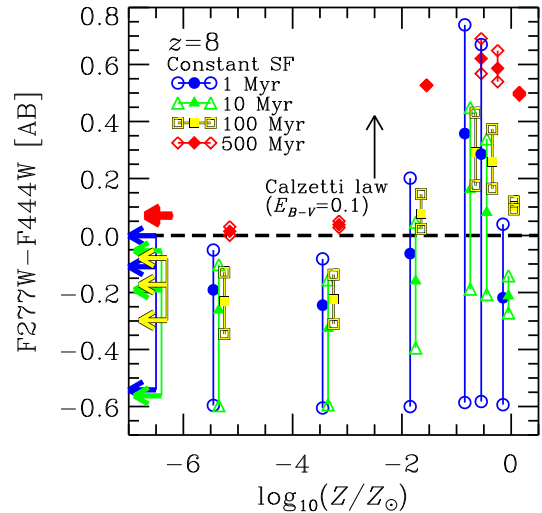}
 \end{center}
 \caption{JWST/NIRCAM F277W$-$F444W colours of $z=8$ galaxies as a
 function of metallicity. The meaning of the symbols are the same as
 Fig.~11. The horizontal dashed line is a criterion to separate the EMP
 cases from other higher metallicity cases.}
\end{figure}

\subsection{Signatures of primordial galaxies}

We discuss what signatures of EMP and Pop III galaxies we can detect in
the near future. First, we present the selection criteria for EMP/Pop
III candidates with broad-band colours. Then, we present more robust
signatures to be searched by follow-up spectroscopy. Table~3 is a
summary of the criteria.

\begin{table}
 \caption[]{Proposed criteria to find extremely metal-poor and
 metal-free galaxies.}
 \setlength{\tabcolsep}{3pt}
 \footnotesize
 \begin{minipage}{\linewidth}
  \begin{tabular}{lc}
   \hline
   Broad-band colours & \\
   \hline
   HST/WFC3 & $J_{125}-H_{160}<-0.15$ for $z\sim7$--8\\
   JWST/NIRCAM & F277W$-$F444W $>0$ for $z\sim6$\\
               & F277W$-$F444W $<0$ for $z\sim8$\\
   \hline
   Rest-frame EW (\AA) & \\
   \hline
   Ly$\alpha$ & $>230$ \\
   H$\alpha$ & $>1900$ \\
   H$\beta$ & $>300$ \\
   He {\sc ii} $\lambda1640$ & $>1$ \\
   $[$O {\sc iii}$]$ $\lambda5007$ & $<20$ \\
   \hline
   Metal/hydrogen ratio & \\
   \hline
   $[$O {\sc iii}$]$ $\lambda5007$/H$\beta$ & $<0.1$ \\
   \hline
  \end{tabular}
 \end{minipage}
\end{table}%

\subsubsection{Broad-band colours}

As discussed with Figure~11, the rest-frame UV colours can be used as
the signature for EMP and Pop III galaxies \citep[see also][]{bou10a}.
However, the colour difference between EMP/Pop III and higher
metallicity cases is relatively small, $\sim0.1$--0.2 mag. Thus, the
observational uncertainty for individual objects prevents us from
distinguishing them, e.g., Figure~12. If future facilities suppress the
uncertainty of the colour, the UV colours will be useful as the
signature for EMP/Pop III stars in individual galaxies. A proposed
criterion is $J_{125}-H_{160}<-0.15$ for $z\sim7$--8.

The discussion with Figure~14 suggests the possibility to use the
rest-frame optical colours as a signature of EMP and Pop III
galaxies. The colour difference between EMP/Pop III and higher
metallicities is larger than those of the UV colours: $\sim0.2$--0.4
mag. In this respect, the optical colours is better than the UV colours.
However, we need space telescopes in order to measure the optical
colours of very high-$z$ galaxies. The JWST and a Japanese project,
Wide-field Imaging Surveyor for High-redshift 
(WISH)\footnote{http://www.wishmission.org/en/index.html} will be
useful. Proposed criteria are F277W$-$F444W $>0$ for $z\sim6$ 
and F277W$-$F444W $<0$ for $z\sim8$.

\subsubsection{EW of emission lines}

If emission lines characterising EMP and Pop III galaxies are strong
enough to be detected, the lines are very useful as the signature of
these primordial galaxies. The He {\sc ii} $\lambda1640$ is the most
discussed feature for Pop III stars \citep[e.g.,][]{sch02,sch03}.
As shown in Figure~7, this line is indeed special feature for the
metal-free case. However, there are two problems to use the line.
One is that the strength of the line is not very strong. The expected EW
is about 1--3 \AA\ in the rest-frame for the star formation duration
longer than 10 Myr. This corresponds to the observed EW of $\la30$ \AA\
for $z<9$. The other is that the strength becomes even weaker if the LyC
escape fraction $f_{\rm esc}>0$. In addition, there is contamination of
the He {\sc ii} line from Wolf-Rayet stars with normal metallicity
\citep{sha06,erb10}. Therefore, we need other signatures to confirm the
galaxies selected by the line to be truly Pop III. In any case, we can
use the line to select the candidates. A proposed criterion is 
EW(He {\sc ii}) $>1$ \AA\ in the rest-frame.

As discussed in section 3.2, EWs of Ly$\alpha$, H$\alpha$, and H$\beta$
can be used as a signature of EMP and Pop III galaxies. However, if we
adopt a lower limit on the EWs to select the primordial galaxies based
on the estimations with $f_{\rm esc}=0$, we will miss some of them which
have $f_{\rm esc}>0$. In addition, Ly$\alpha$ photons undergoes resonant
scattering in the ISM, and sometimes, the EW with higher metallicities
is boosted by dust \citep{neu91}, and may exceed the criterion and
become contaminant. There is no such case with H$\alpha$ and
H$\beta$. Thus, these lines are more reliable signatures. Proposed
criteria are EW(Ly$\alpha$) $>230$ \AA, EW(H$\alpha$) $>1900$ \AA, and
EW(H$\beta$) $>300$ \AA\ in the rest-frame, or to avoid very young
($\sim1$ Myr) galaxies with higher metallicities, EW(Ly$\alpha$) $>460$
\AA, EW(H$\alpha$) $>3200$ \AA, and EW(H$\beta$) $>540$ \AA\ in the
rest-frame.

With the EW of [O {\sc iii}] $\lambda$5007, we can select all the
primordial galaxies by an upper limit on the EW. However, in this case,
we will have contaminants of galaxies with higher metallicities and 
$f_{\rm esc}>0$. A proposed criterion is EW(O {\sc iii}) $<20$ \AA\ in
the rest-frame.

\subsubsection{[O {\sc iii}]/H$\beta$ line ratio}

Metal-to-hydrogen line ratio is free from the uncertainty of 
$f_{\rm esc}$, unlike the EWs. As found from Figure~1, the ratio of 
[O {\sc iii}] $\lambda5007$/H$\beta$ $<0.1$ if $\log_{10}(Z/Z_\odot)<-3$.
Although the ratio also becomes $<0.1$ if $\log_{10}(Z/Z_\odot)>0.6$
based on the empirical relation reported by \cite{nag06}, such very high
metallicity galaxies would not exist at very high-$z$, or if they exist,
itself is very interesting. Therefore, searching a galaxy with the ratio
$<0.1$ is very attractive.

Based on Figure~2, the H$\beta$ luminosity per a unit star formation
rate of EMP/Pop III galaxies is $\sim20$--$30\times10^{40}$ erg s$^{-1}$
($M_\odot$ yr$^{-1}$)$^{-1}$ if the star formation duration is longer 
than 10 Myr and $f_{\rm esc}=0$. Thus, the expected upper limit on the
[O {\sc iii}] luminosity of EMP/Pop III galaxies is
$<2$--$3\times10^{40}$ erg s$^{-1}$ ($M_\odot$ yr$^{-1}$)$^{-1}$. This
corresponds to the [O {\sc iii}] line flux $<3\times10^{-20}$ erg
s$^{-1}$ cm$^{-2}$ ($M_\odot$ yr$^{-1}$)$^{-1}$ for a $z=8$ source. The
expected line sensitivity of the Near Infrared Spectrograph (NIRSpec) on
the JWST is $3\times10^{-19}$ erg s$^{-1}$ cm$^{-2}$ at 4.5 $\mu$m at
S/N=10 with R=1000 mode in 100,000 s exposure 
time.\footnote{http://www.stsci.edu/jwst/instruments/nirspec/sensitivity/R1000{\_}line.pdf}
Therefore, we can reach [O {\sc iii}]/H$\beta$ $<0.1$ at 3-$\sigma$ for a
10 (or 3) $M_\odot$ yr$^{-1}$ galaxy at $z=8$ in 10,000 (100,000) s with
the JWST/NIRSpec.

\subsubsection{Balmer jump}

As discussed in previous sections, we expect a significant Balmer jump
in spectra of EMP/Pop III galaxies. Let us examine if we detect this
feature with spectroscopy. Based on Figure~3 (a), we expect the continuum
level around the Balmer jump (3646 \AA\ in the rest-frame) to be (2--10)
$\times 10^{27}$ erg s$^{-1}$ Hz$^{-1}$ ($M_\odot$ yr$^{-1}$)$^{-1}$,
depending on $f_{\rm esc}$ and the star formation duration. Thus, we
expect an observed flux density of (2--10) nJy ($M_\odot$
yr$^{-1}$)$^{-1}$ for $z=8$ galaxies. On the other hand, the
expected sensitivity of the JWST/NIRSpec is about 200 nJy at 3--4 $\mu$m
at S/N $=10$ with R$=$100 mode in 10,000 s exposure 
time.\footnote{http://www.stsci.edu/jwst/instruments/nirspec/sensitivity/R100{\_}cont.pdf}
Therefore, we can detect continuum of a 10 $M_\odot$ yr$^{-1}$ galaxy at
$z=8$ in 100,000 s exposure with very high significance of S/N $=3$--15.
Namely, we can detect the Balmer jump of primordial galaxies at $z=8$
with the JWST/NIRSpec.

\section{Summary}

This paper presents a spectral model
from UV to optical in the rest-frame of galaxies with various
metallicities. A special feature of the model is the nebular emission of
lines and continuum. We have calculated intensities of 119 emission
lines from H Ly$\alpha$ to 1 $\mu$m in the rest-frame with the public
photo-ionization code {\sc cloudy} 08.00 \citep{fer98}. The stellar
spectra input into the code are generated by {\sc starburst99}
\citep{lei99} for metallicity $Z\ge1/50Z_\odot$ or taken from
\cite{sch02,sch03} for EMP and metal-free cases. We input the stellar
spectrum with the same metallicity as that in the nebular gas into the
photo-ionization code. After exploring a wide range of nebular parameters
(Table~1) which are appropriate to real H {\sc ii} regions, we have
derived average intensities of 119 emission lines relative to H$\beta$
(Figure~1 for the 6 strongest metal lines) and temperatures of the
nebular gas (Table~2) as a function of metallicity. The emission line
intensities are presented in Appendix as a machine-readable form.

The resultant spectra of galaxies show some interesting features in the
rest-frame optical (Figure~3). In the EMP and metal-free cases, we
find a strong Balmer jump, which is a spectral plummet towards longer
wavelength, by the nebular bound-free continuum (Figures~4 and 5). As a
result, broad-band colours straddling the Balmer jump becomes
bluer than the stellar one (see Figure~13). However, strong Balmer
emission lines can fill in the jump and can make the colours even redder
than the stellar one (Figure~13). In higher metallicity cases, some
emission lines of oxygen like [O {\sc iii}] are so strong that they
redden the broad-band colours \citep[Figure~13; see also e.g.,][]{sch09}.

We have extensively discussed the signatures of EMP and metal-free
galaxies expected from the model. In current and future observational
data, we can obtain a sample of galaxies at high-$z$ by the standard
drop-out technique. If we can select EMP and metal-free candidates from
it by broad-band data, it is very useful to select the target for
follow-up spectroscopy. The bluest galaxies in rest-frame UV are a good
target (Figures~11 and 12). An example criterion in the colour with the
HST/WFC3 is $J_{125}-H_{160}<-0.15$ for $z\sim7$--8. The blue galaxies
with the UV spectral slope $\beta=-3\pm0.2$ reported by \cite{bou10a}
are likely to be EMP or metal-free although we cannot deny the
possibility that these galaxies have normal metallicity if they are very
young ($\sim1$ Myr) and have a large LyC escape fraction  ($f_{\rm
esc}>0.5$) (Figure~11). Rest-frame optical colours are also
useful because there are some metallicity indicators such as the strong
Balmer jump, Balmer series lines, and oxygen lines (Figures~13 and 14).
For example, the criteria with the JWST/NIRCAM colour are F277W$-$F444W
$>0$ for $z\sim6$ and F277W$-$F444W $<0$ for $z\sim8$.

After selected the candidates, we will perform follow-up spectroscopy
for them, and then, measure the metallicity. At this point, equivalent
widths (EWs) of emission lines become useful as an indicator. We have
examined EWs of Ly$\alpha$, He {\sc ii} $\lambda1640$, H$\alpha$,
H$\beta$, and [O {\sc iii}] $\lambda5007$ as a function of metallicity,
star formation duration, and LyC escape fraction (Figures 6--10). For
the duration $>10$ Myr and $f_{\rm esc}=0$, the EMP and metal-free
criteria of the rest-frame EW are as follows: EW(Ly$\alpha$) $>230$ \AA,
EW(He {\sc ii}) $>1$ \AA, EW(H$\alpha$) $>1900$ \AA, EW(H$\beta$) $>300$
\AA, and EW(O {\sc iii}) $<20$ \AA. Note that the Ly$\alpha$ and He {\sc
ii} criteria have a further uncertainty due to the resonant scattering
in the ISM and IGM and due to the contribution of Wolf-Rayet stars,
respectively. A criterion independent of the star formation duration and
$f_{\rm esc}$ is the metal-to-hydrogen line ratio. The most easily
detectable ratio is [O {\sc iii}] $\lambda5007$/H$\beta$. If we find the
ratio $<0.1$ from a galaxy, it is EMP or metal-free. This ratio at
$z\sim8$ can be detectable by spectroscopy with the JWST/NIRSpec within
a reasonable exposure time. In addition, we find that the Balmer jump
can be also detected with the JWST/NIRSpec spectroscopy.

\section*{Acknowledgments}

The author thanks to Roberto Maiolino, reviewer, for understanding this
paper well and giving some useful comments. The author also thanks to
Ikuru Iwata, Toru Yamada, and Koji Ohta for discussions and suggestions
which let him initiate this work. The author is supported by the
Institute for Industrial Research, Osaka Sangyo University and by
KAKENHI (the Grant-in-Aid for Young Scientists B: 19740108) by The
Ministry of Education, Culture, Sports, Science and Technology (MEXT) of
Japan.

\section*{Appendix}

A machine-readable tables of emission line intensities relative to
H$\beta$; Table~A1 is the no dust model and Table~A2 is the dusty model.

\label{lastpage}

\end{document}